\documentclass[aps,prl,twocolumn,amsmath,amssymb,showpacs,
floatfix]{revtex4}

\usepackage{epsfig}
\usepackage{bm}
\begin{document}


\title{
New method for the 3D Ising model
}

\author{
S.G. Chung}
\email[]{sung.chung@wmich.edu}
\affiliation{
Department of Physics and Nanotechnology Research and Computation Center,
 Western Michigan University, Kalamazoo, MI 49008-5252, USA}

\date{\today}

\begin{abstract}
A simple, general and practically exact method is developed for 
the equilibrium properties of 
the macroscopic physical systems with translational symmetry. Applied to the Ising model 
in two and three dimension, a modest calculation gives the spontaneous magnetization 
and the specific heat to less than 1\% error.
\end{abstract}

\pacs{05.50.+q, 64.60.Cn, 75.10.Hk}
\keywords{D=2,3 dimensional Ising model, novel many-body formalism
, numerically exact method, spontaneous magnetization}

\maketitle

To calculate the equilibrium properties of a given physical system will probably be the most 
basic theoretical task.  In 1944, Onsager \cite{onsager} calculated the partition function of the two 
dimensional Ising model analytically and demonstrated the power of the exact solution, 
followed by the exact spontaneous magnetization 
by Yang \cite{yang}.  Since then some remarkable progresses have been made in 
two-dimensional classical and one-dimensional quantum systems analytically \cite{mcc,bax,
and,chu}, and by 
numerical methods such as DMRG (density matrix renormalization group) \cite{dmrg}.  
However, reflecting the three-dimensionality of our world, much needed knowledge is 
for three dimension, and several thousand papers have been said written on the three 
dimensional Ising model.  Among others, a tremendous effort has been paid to the 
precise determination of the critical temperature $T_c$ in association with the concept 
of universality near the critical point \cite{fis}.  Owing to the concerted effort of 
RG-finite size scaling-Monte Carlo (MC) method, $T_c$ is known 
as precise as 4.5114(1) \cite{lan}, as well as the spontaneous magnetization 
\cite{tal}.  
The effort of 
high and low temperature expansions by the diagrammatic method for 
$T_c$ is also remarkable \cite{but}. 
Yet another noteworthy recent progress is along the line of 
DMRG \cite{gen}. 

We here present a new method 
for the equilibrium properties of the 
macroscopic physical systems with translational symmetry.   
Our method is {\it purely algebraic}, works directly on a 
$\infty~{\rm x}~ \infty~{\rm x}~ \infty$ lattice and, other than seeking 
a convergence in entanglement 
space (see below), it does not invoke any other notions such as 
numerical RG, nor make any approximations.
We have calculated the spontaneous 
magnetization and the specific heat in two and three dimensional Ising models to less than 
1\% error.  Note that without the help of RG, the 1\% precision
is what one can typically expect from exact methods such as Bethe Ansatz \cite{and,chu}.
The new method is simple, general and computationally efficient.  In fact,
 our 30 minutes calculation already gives fairly precise results as demonstrated below.

Let us consider the ferromagnetic Ising model on a square or simple cubic lattice,
\begin{equation} \label{eq1}
H=-\sum_{(i,j)}\sigma_i\sigma_j
\end{equation}
where $\sigma_i$ takes + or - 1 and the summation is over the nearest neighbor pairs.
The partition function is given by
$Z={\rm Tr}[{\rm exp}(-\beta H)]$, where ${\rm Tr}$ means to take trace over the $2^N$ 
spin configurations with N being the number of lattice sites and $\beta=1/kT$.  
We follow the following steps for the two dimensional case.

{\it First}, note that the local pair-density matrix can be written as  
${\rm exp}(\beta \sigma_i \sigma_j)={\rm cosh}(\beta)+{\rm sinh}(\beta)
\sigma_i \sigma_j$.  This is the simplest case
of more general statement that any local density matrix
 when regarded as a real symmetric matrix can be written by singular value decomposition (SVD) as
$A_{ij}=\sum_k{V_{ik}\lambda_k V_{jk}}$, where
$V_k$ and $\lambda_k$ are eigenvectors and eigenvalues
of the local density matrix $A$.  Denoting these two terms by a "bond index" $i=1, 2$, the 
partition function is written as a summation over all the bond index configurations
on the square lattice.  Now each term in this summation is a product over all the sites of
a local spin operator constructed from 
$\lambda_1$ and $\lambda_2\sigma_i$, 
depending on the four bond indexes surrounding the i-th spin, where 
$\lambda_1=\sqrt{{\rm cosh}(\beta)}$ and $\lambda_2=\sqrt{{\rm sinh}(\beta)}$.
For example at site $i$, if the surrounding four bond indexes are all 1, then
the local spin operator at the site is $\lambda_1^4$ and its trace is, 
$Tr\lambda_1^4=2\lambda_1^4$. If one of the four bond indexes is replaced by 2,
then the local spin trace is, $Tr\lambda_1^3\lambda_2\sigma_i=0$. Let us denote
these local spin traces as $\Gamma_{ijkl}$ with the four bond indexes 
$i,j,k$ and $l$ taking 1 or 2. Clearly, 
$\Gamma_{ijkl}=2\lambda_1^4,2\lambda_1^2\lambda_2^2,2\lambda_2^4$
for $i+j+k+l$ = 4,6, and 8 respectively and zero otherwise.
Performing the spin trace over all the sites, we have
\begin{equation} \label{eq2}
Z=\sum_{bond}\Pi_{site}\Gamma_{ijkl}
\end{equation}
{\it Second}, by virtue of infinite system, we can conveniently assume the periodic 
boundary condition in both horizontal and vertical directions. 
Let us denote the leftmost horizontal bond indexes by a vector ${\bf a_1}$ and the rightmost 
ones by ${\bf a_{N+1}}$. The periodic boundary condition is then expressed by the Cronecker delta
$\delta_{{\bf a_1}{\bf a_{N+1}}}$ which can be expanded by an orthonormal 
complete set $\{\varphi_n({\bf a})\}$ as 
$\delta_{{\bf a_1}{\bf a_{N+1}}}=\sum_n\varphi_n({\bf a_1})\varphi_n({\bf a_{N+1}})$.  
Note that all the quantities in this paper are real.  Let us now denote the rightmost 
column composed of a vertical one-dimensional array of $\Gamma$s 
as $K({\bf a_N},{\bf a_{N+1}})$.  
Note that we can certainly regard this quantity as a function of the two bond-index 
vectors ${\bf a_N}$ and ${\bf a_{N+1}}$  
after summing over the vertical bond indexes associated with this column.  Now choose 
the above introduced eigenstate as the one which satisfies a transfer-matrix 
eigenvalue equation
\begin{equation} \label{eq3}
K({\bf a_N},{\bf a_{N+1}})\varphi_n({\bf a_{N+1}})=\mu_n\varphi_n({\bf a_N})
\end{equation}
where and below the repeated indexes imply a summation. 
Repeatedly using Eq(\ref{eq3}), the partition function is evaluated 
as $Z=\sum_n\mu_n^N \to \mu_0^N$ 
in the large N limit, where the suffix 0 indicates the largest eigenvalue.
The specific heat can be obtained from the free energy $F=-kT\ell n(Z)$ by differentiating twice with 
respect to temperature. To avoid unnecessary error due to numerical differentiation, 
however, one might calculate the internal energy per site, $E=-D<\sigma_i\sigma_j>$ where D 
is the space dimensionality and $<\sigma_i\sigma_j>$ means the correlation of nearest neighbor spins, 
and then the specific heat is obtained by differentiating the energy with respect 
to temperature once.  Likewise the spontaneous magnetization per site is given by 
$M=<\sigma_i>$.
{\it Third}, we write $\varphi_0({\bf a_{N+1}})$ in the same form as the operator 
$K({\bf a_N},{\bf a_{N+1}})$ with each $\Gamma_{ijkl}$ being replaced by a three index 
quantity $\zeta_{lmn}$, 
thus without the index ${\bf a_N}$. This is justified as a Schmidt decomposition
\cite{nie} of eigen functions with translational invariance. Or more simply, one can derive this form
by a successive use of SVD.  Consider, for example, a wave function $\varphi(a_1a_2a_3a_4)$. 
Regarding this as a matrix of the left index $a_1$ and the right index $\{a_2a_3a_4\}$, we have
$\varphi(a_1a_2a_3a_4)=\sum_\alpha{A_{a_1\alpha}\rho_\alpha 
B_{\{a_2a_3a_4\}\alpha}}$.  The quantity $B$ can in turn be regarded as a matrix
of the left index $\{a_2 \alpha\}$ and the right index $\{a_3a_4\}$, thus SVD gives
$B_{\{a_2a_3a_4\} \alpha} = \sum_\beta{C_{\{a_2 \alpha\}\beta} 
\lambda_\beta D_{\{a_3a_4\} \beta}}$.  Likewise, 
$D_{\{a_3a_4\} \beta}=\sum_\gamma{E_{\{a_3 \beta\}\gamma} \Delta_\gamma F_{a_4\gamma}}$.
Putting together, rewriting $A_{a_1 \alpha}$ as $A_\alpha(a_1)$, 
$C_{\{a_2 \alpha\} \beta}$ as $C_{\alpha\beta}(a_2)$, 
$E_{\{a_3 \beta\} \gamma}$ as $E_{\beta\gamma}(a_3)$ and
$F_{a_4 \gamma}$ as $F_\gamma(a_4)$, and appropriately absorbing $\rho_\alpha$,
$\lambda_\beta$ and $\Delta_\gamma$ into the matrixes $A$, $C$, $E$ and $F$, one gets
$\varphi(a_1a_2a_3a_4)=A_\alpha(a_1)C_{\alpha\beta}(a_2)E_{\beta\gamma}(a_3)F_\gamma(a_4)$.
For an infinite system with translational symmetry and periodic boundary condition, one
arrives at the claimed form, with $\alpha$, $\beta$ and $a_{i-1}$, e.g., corresponding to 
$l$, $m$, and $n$ in the above $\zeta_{lmn}$,
\begin{equation} \label{eq4}
\varphi_0({\bf a_{N+1}})=...\zeta_{\alpha\beta}(a_{i-1})\zeta_{\beta\gamma}(a_i)\zeta_{
\gamma\delta}(a_{i+1})...
\end{equation}
Fig.1 represents the transfer matrix eigenvalue equation Eq(\ref{eq3}) schematically.
In quantum information theory, these indexes $\alpha$, $\beta$ and $\gamma$  are known as {\it entanglement} \cite{nie}.
Considering only 1 for these indexes is a simple 
mean-field-like approximation for $\varphi_0({\bf a_{N+1}})$. Allowing larger values, one takes into account the effect of correlation 
with increasing precision.  
An important note here is that Eq(\ref{eq4}) is not peculiar to the Ising models, but
rather a general statement for any macroscopic systems with translational symmetry. 
Another important note is, as is evident from the derivation above, the range of values
which the entanglements $\alpha$, $\beta$ and $\gamma$ can take, grows exponentially for 
increasingly large systems.  Thus the success of the present method crucially relies on
a rapid convergence of physical quantities with the increase of entanglement.   
{\it Fourth}, we handle the eigenvalue problem (3) as 
a variational problem, namely we maximize 
the quantity $\mu_0=\varphi_0 K \varphi_0/\varphi_0\varphi_0$ by iteration starting with an input 
state for $\varphi_0$. First consider the numerator.  
\begin{figure}
\label{fig1}
\epsfig{file=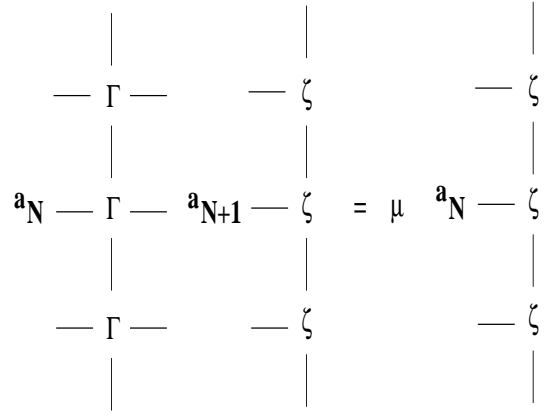,width=7.0cm,height=5.5cm}
\caption{
Schematic figure of the transfer matrix eigenvalue equation, Eq(\ref{eq3}).
}
\end{figure}
A local ingredient of this quantity 
is $A_{ll'e,mm'f}\equiv\zeta_{lmn}\Gamma_{nefn'}\zeta_{l'm'n'}$. 
Since A is real symmetric, we can use SVD again and write 
as $A=X\nu X^{tr}$ where the matrixes $X$ and $\nu$ are made up of eigenvectors and 
eigenvalues of $A$ 
and $tr$ means the transpose.  Since $X$ is orthogonal, the 
summation over the combined entanglement-bond indexes $ll'e$ in the numerator 
can be done for N-1 times, and thus we only keep the largest eigenvalue $\nu_0$ 
and eigenstate ${\bf x_0}$. We have $\varphi_0 K\varphi_0=\nu_0^{N-1}{\bf x_0}^{tr}
A{\bf x_0}$. The denominator can be handled likewise.  Let us denote the corresponding 
largest eigenvalue and eigenstate as $\rho_0$ and ${\bf y_0}$.  Note that $\mu_0$ now 
contanins $\zeta_{lmn}$ in quadratic form.  Maximizing this quantity with respect to
$\zeta_{lmn}$  then leads to a generalized eigenvalue problem:
\begin{equation} \label{eq5}
x_{0ll'e}\Gamma_{enn'f}x_{0mm'f}\zeta_{l'm'n'}=\tilde{\mu}_0 y_{0ll'}\delta_{nn'}
y_{0mm'}\zeta_{l'm'n'}
\end{equation}
where $\mu_0=\tilde{\mu}_0\nu_0^{N-1}/\rho_0^{N-1}$. 
We solve Eq(5) for the next $\zeta$ and continue until convergence.
{\it Finally} after the convergence, we can calculate the internal energy and 
spontaneous magnetization in terms of the obtained eigenvalues and eigenstates.  
The procedure is the same as above with only difference being
 one (two) of $\Gamma$s in the expression for the partition function should be replaced
by $\tilde{\Gamma}_{ijkl}=2\lambda_1^3\lambda_2,2\lambda_1\lambda_2^3$ 
for $i+j+k+l$ = 5, 7 respectively and zero otherwise for calculating $<\sigma_i>$ 
($<\sigma_i\sigma_j>$). We have,
\begin{equation} \label{eq6}
M={\bf x_0}^{tr}\tilde{A}{\bf x_0}/(\tilde{\mu}_0\rho_0)
\end{equation}
where $\tilde{A}$ is the same as $A$ with $\Gamma$ replaced by $\tilde{\Gamma}$. 
Likewise
\begin{equation} \label{eq7}
E=-D{\bf x_0}^{tr}\tilde{A}^2{\bf x_0}/(\tilde{\mu}_0\rho_0\nu_0)
\end{equation}

The extension to three dimension is straightforward. $\Gamma$ and $\tilde{\Gamma}$ 
now have 6 indexes, $\Gamma_{ijklmn}=2\lambda_1^6,2\lambda_1^4\lambda_2^2,
2\lambda_1^2\lambda_2^4,2\lambda_2^6$  for $i+j+k+l+m+n$ = 6,8,10 and 12 respectively
 and zero otherwise and $\tilde{\Gamma}_{ijklmn}=2\lambda_1^5\lambda_2,
2\lambda_1^3\lambda_2^3,2\lambda_1\lambda_2^5$ for $i+j+k+l+m+n$ = 7,9 and 11 
respectively and zero otherwise, and the bond index vector ${\bf a}$ represents a two dimensional array.  
The largest eigenstate $\varphi_0({\bf a})$ of the operator $K$ is now expressed 
as a product of $\zeta_{abcd\alpha}$, the first four indexes denoting entanglement 
while the last one a bond index. 
For notational simplicity, 
the suffix zero for denoting the largest eigenvalue and eigenstate will be dropped from now on.
The maximization of $\mu=\varphi K\varphi/\varphi\varphi$  
 now takes the structure of a "Russian dole". 
  In fact the calculation of the numerator is identical with that of the 
partition function in two dimension with $\Gamma$ 
replaced by $B=\zeta_{abcd\alpha}\Gamma_{\alpha efgh \beta}\zeta_{ijkl\beta}$.  
The bond indexes are 
replaced by the composite of bond-entanglement indexes (aei), (bfj) etc. which we will 
denote by a single index below when possible.  $B$ can then be expressed like $B_{ijkl}$. 
We consider an eigenvalue problem similar to Eq(\ref{eq3}).  Let us write the 
largest eigenvalue as $\Delta$ and the local constituent of the eigenstate as 
$\eta_{tt'l}$ (corresponds to $\zeta_{lmn}$ in two dimension) where the 
first two indexes represent 
the new entanglement indexes and the last a composite of bond-entanglement indexes.
  Let us denote the quantities corresponding to $\nu_0$, ${\bf x_0}$ and $\rho_0$ 
in two dimension as $\tau$, ${\bf u}$ and $\epsilon$. The denominator is likewise treated.  
Here $\Gamma$ is replaced by $C=\zeta_{abcd\alpha}\zeta_{ijkl\alpha}$, 
and the composite of bond-entanglement indexes is now (ai) etc.  Corresponding to 
$\Delta,\eta,\tau,{\bf u}$ and $\epsilon$, we write $\Omega,\xi,\tau_d,{\bf v}$ and $\epsilon_d$.
It is convenient to define a quantity $G_{pti,p't'l} \equiv \eta_{pp'j}B_{ijkl}\eta_{tt'k}$ and 
associated $\tilde{G}$ which is obtained from $G$ with $\Gamma$ replaced by $\tilde{\Gamma}$. 
Likewise define $O_{pti,p't'l} \equiv \xi_{pp'j}C_{ijkl}\xi_{tt'k}$.  With these notations, we have
\begin{figure}
\label{fig2}
\epsfig{file=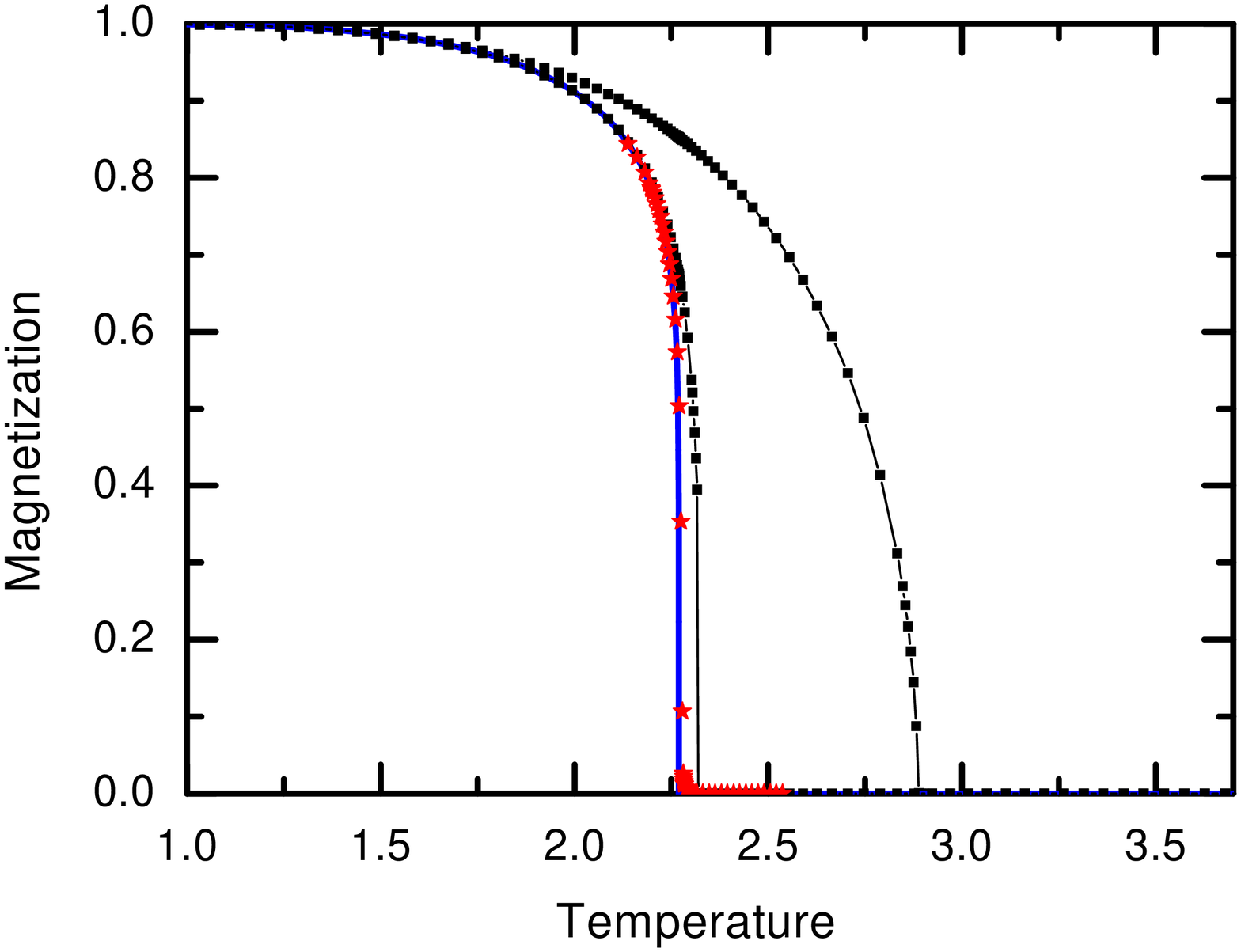,width=9cm,height=7.5cm}
\caption{
Spontaneous magnetization vs. temperature in two dimension.  
From right to left, the entanglement p=1, 2 and 8 (star, red online).  
The leftmost solid line (blue online) is the exact result of Yang \cite{yang}.
}
\end{figure}
\begin{equation} \label{eq8}
\mu=W{\bf u}^{tr}G{\bf u}/{\bf v}^{tr}O{\bf v}
\end{equation}
with $W=\Delta^{N-1}\tau^{N-1}\epsilon_d^N/(\epsilon^N\Omega^{N-1}\tau_d^{N-1})$. 
Note that $\zeta$ is quadratically involved here, and therefore the maximization 
of $\mu$ gives a similar generalized eigenvalue problem as Eq(\ref{eq5}),
\begin{eqnarray} \label{eq9}
u_{ptqsr}\eta_{pp'abc}\Gamma_{\alpha rr'cc'\beta}\eta_{tt'a'b'c'}u_{p't'q's'r'}
\zeta_{sbs'b'\beta}
\nonumber\\
=\tilde{\mu}v_{ptqs}\xi_{pp'ab}\delta_{\alpha\beta}
\xi_{tt'a'b'}v_{p't'q's'}\zeta_{sbs'b'\beta}
\end{eqnarray}
where $\mu=\tilde{\mu}W$.  When the iteration is converged, the spontaneous magnetization 
and the internal energy in three dimension are calculated as,
 note the same formula Eq(\ref{eq6}) and Eq(\ref{eq7}) in two dimension,
\begin{equation} \label{eq10}
M={\bf u}^{tr}\tilde{G}{\bf u}/(\tilde{\mu}\tau_d)
\end{equation}
\begin{equation} \label{eq11}
E=-D{\bf u}^{tr}\tilde{G}^2{\bf u}/(\tilde{\mu}\tau\tau_d)
\end{equation}
\begin{figure}
\label{fig3}
\epsfig{file=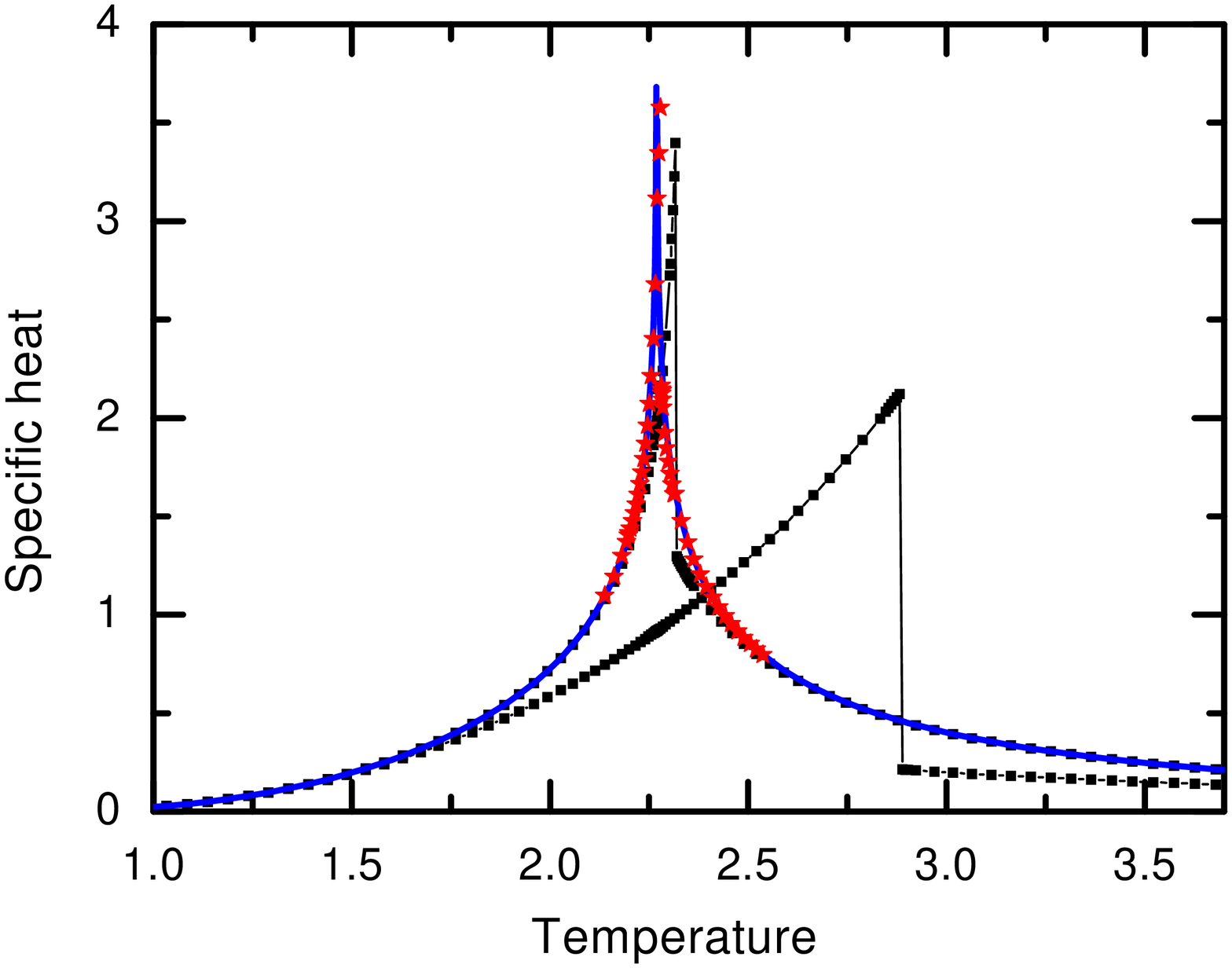,width=9cm,height=7.5cm}
\caption{
The same as Fig.2 for the specific heat. The peak is at $T=2.278$. 
The thick solid line (blue online) is the exact 
result of Onsager \cite{onsager}.
}
\end{figure}

The results for the spontaneous magnetization and the specific heat in two dimension are 
shown in Fig.2 and 3.  
The results with the entanglement p = 8 are indistinguishable from 
the exact ones of Onsager and Yang.  The obtained critical temperature 2.278 is 0.4\% off 
the exact value 2.269185.  The results in three dimension are shown in Fig.4 and 5. 
\begin{figure}
\label{fig4}
\epsfig{file=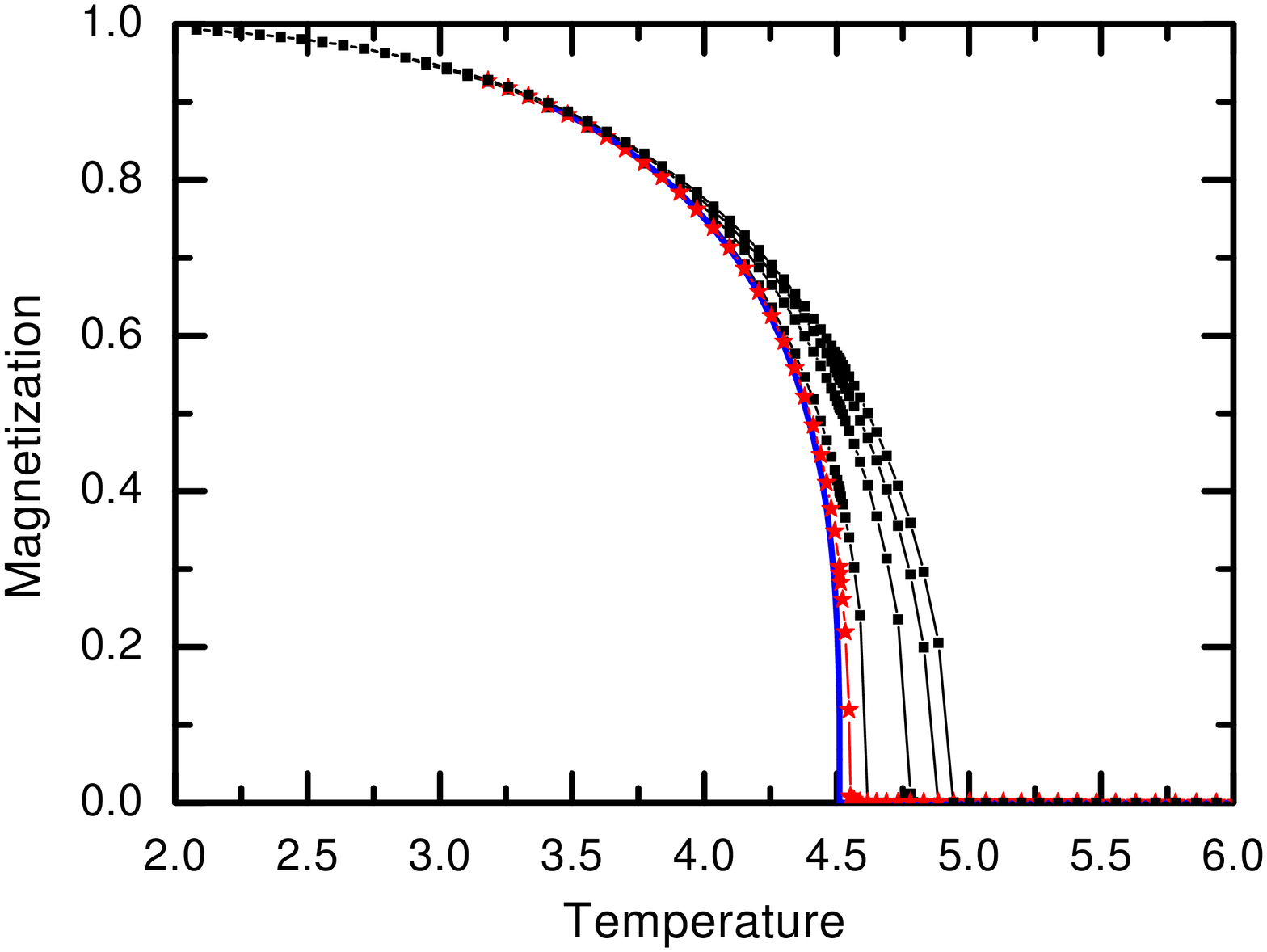,width=9cm,height=7.5cm}
\caption{
Spontaneous magnetization vs. temperature in three dimension. From right to left, the entanglement 
(p,q)=(1,1), (1,2), (2,1), (2,2) and (2,4) (star, red online), where p 
refers to the entanglement associated with $\zeta$ and q with $\xi$ and $\eta$.
For p=1, the (1,2) result is the converged one.
 The leftmost solid line (blue online) is the empirical 
formula from the RG-finite-size-scaling-MC method \cite{tal}. 
} 
\end{figure}
\begin{figure}
\label{fig5}
\epsfig{file=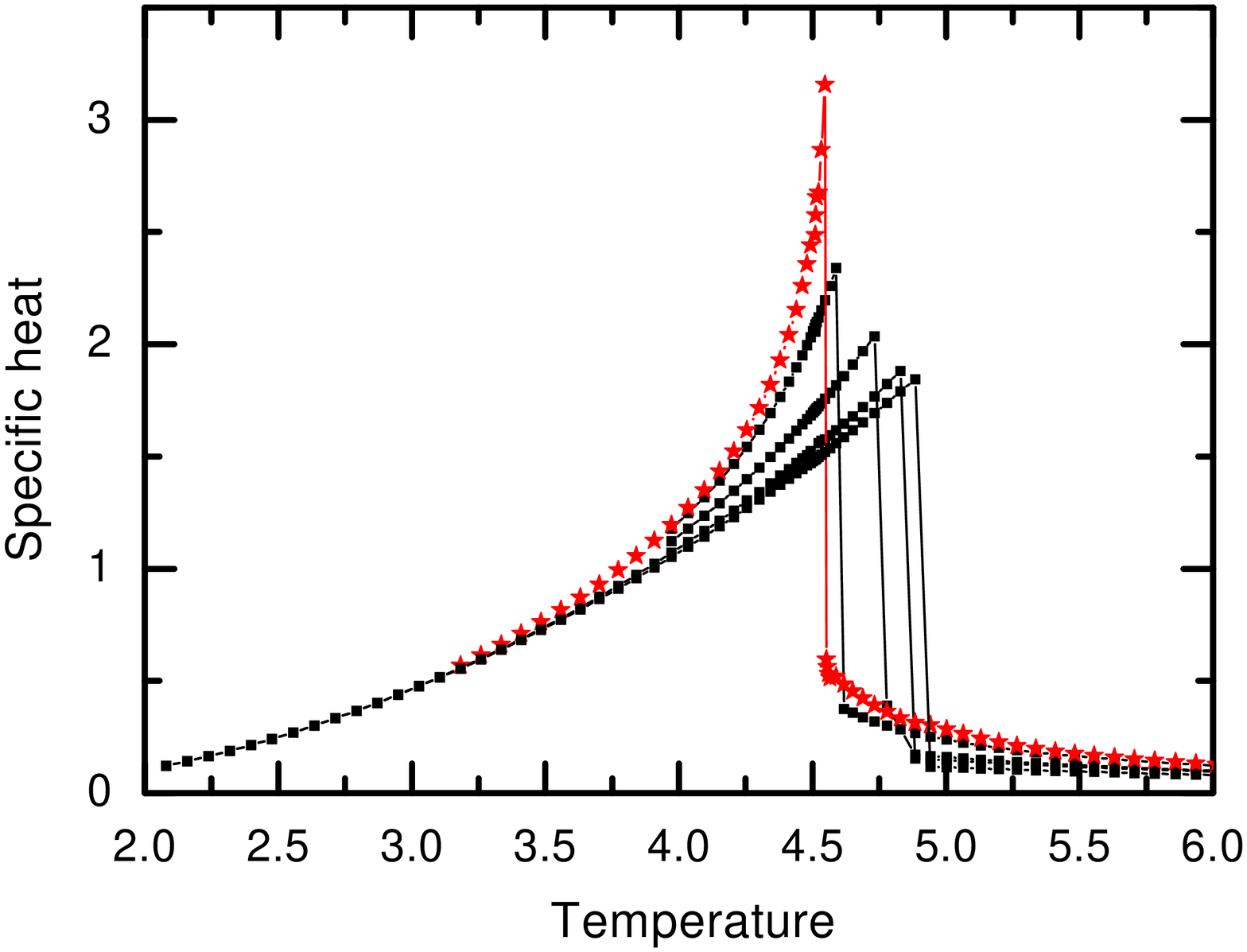,width=9cm,height=7.5cm}
\caption{
The same as Fig.4 for the specific heat. The peak is at T=4.547.}
\end{figure}
The result
for the spontaneous magnetization with the entanglement (p,q) = (2,4) is very close to 
the empirical formula due to the RG-finite-size-scaling-MC method \cite{tal}, 
\begin{eqnarray} \label{eq12}
M(t)=-t^{0.32694109}(1.6919045
\nonumber\\
-0.34357731t^{0.50842026}-0.42572366t),
\end{eqnarray}
for $0.0005<t<0.26$ where t=1-0.2216544$kT$. The obtained critical temperature 4.547 is 0.8\% off the believed 
exact value 4.5114(1).  The specific heat in three dimension has a marked difference from 
two dimension, namely it is highly asymmetric near the critical point.  In fact our 
result with the entanglement (2,4) has quite a resemblance to both the MC 
prediction \cite{bin} (also on page 479 of \cite{itz}) 
and the specific heat experiment on argon at the 
gas-liquid phase transition \cite{bag} (also on page 13 of \cite{fis}).
It may be worth pointing out that the method can calculate equilibrium properties 
 easily to a high precision.  As is seen in Fig.1-4, 
the p=2 case in two dimension and the (2,2) case in three dimension give fairly 
precise results, but it took only 30 minutes using 
a single PC of about 1GHz processing speed.

Because of simplicity and 
generality of the method, many 
applications and further progresses are anticipated, 
but from a theoretical viewpoint, the entanglement p = 3 case 
in three dimension needs to be resolved first to see the swift convergence with the increase 
of entanglement. 
The simplest case (3,1), however, was done easily giving at most (near the transition point) 
$10^{-4}$ relative correction to the case (2,1), indicating that 
the correction from p = 3 will indeed be very small. After confirming this point, the method 
would serve as an exact method for most of the  purposes in equilibrium statistical physics.  
As for a much more accurate analysis of the transition point such as $T_c$ and
various critical exponents beyond 1\% level, as was done in \cite{lan,tal,but,gen2}, 
it is interesting to see how a combination of RG and the present method performs.  

In conclusion, the essence of the new method shall be summarized and further clarified.  First, it is {\it not}
a kind of cluster mean field theories or transfer-matrix mean field theories as 
used, e.g., in the coherent anomaly method \cite{mas}.  In our word, the cluster mean field theories
are described as the case with entanglement $=1$, but each $\zeta$ there represents a cluster of bonds.  
By considering entanglement greater than 1, we go beyond mean field theories.  
Second, we have not made any assumptions, nor used any peculiar procedures
only applicable to the Ising models.  We have used SVD repeatedly and the key idea of our
method is to {\it fully implement the translational symmetry}.  Third, one might still press 
for an evidence of power of the new method.  For that matter, we simply note that the new method 
has been successfully extended to the 1D Hubbard model, 
reproducing the main results of Bethe Ansatz \cite{chu2}. 
Finally, a deep question is if our method offers a new physical concept or picture.
For this, we point out that the state $\varphi_0$ for the Hubbard model includes, as a 
special case, the resonating-valence-bond (RVB) state proposed by Anderson for the 
high-temperature superconductors \cite{and2}.  We note that the {\it resonating} character of the
Anderson wave function is a natural consequence of the translational symmetry.  In this sense,
our state $\varphi_0$ is {\it always resonating}.  It could be any states, superconducting,
insulating, matallic or magnetic.  Our method thus opens a possibility of 
precise determination of the phase diagram of the 2D Hubbard model. An effort toward this goal 
is currently under way, and will be reported in a future publication.

\begin{acknowledgments}
I thank Yasutami Takada for his critical reading of the manuscript and valuable comments.  
This work was partially supported by the NSF under grant No. PHY060010N
and utilized the IBM P690 at the National Center for Supercomputing Applications
at the University of Illinois at Urbana-Champaign. 
A part of this work was done while I was a visitor at the 
Max Planck Institut Physik der komplexer Systeme in Dresden, Germany in June, 2005.  
I thank their warm hospitality.
\end{acknowledgments}

\bibliography{anm}

\end{document}